\documentclass[9pt,twocolumn,twoside]{osajnl}
	\usepackage{flushend}
	
	\journal{ao} 
	
	\setboolean{shortarticle}{false} 
	
	\title{3D tracking the Brownian motion of colloidal particles using digital holographic microscopy and joint reconstruction}
	
	\author[1,*]{Nicolas Verrier}
	\author[1]{Corinne Fournier}
	\author[1]{Thierry Fournel}
	
	\affil[1]{Laboratoire Hubert Curien - UMR 5516-CNRS-Universit\'e Jean Monnet- 18 Rue du Professeur Beno\^it Lauras 42000 Saint-Etienne, France}
	
	\affil[*]{Corresponding author: nicolas.verrier@univ-st-etienne.fr}
	
	\dates{Compiled \today}
	
	\ociscodes{(090.1995) Holography: Digital holography, (100.3190) Image processing: Inverse problems, (100.6640) Image processing: Superresolution; (100.3010) Image processing: Image reconstruction techniques,  (120.3940) Instrumentation, measurement, and metrology: Metrology}
	
	\doi{\url{http://dx.doi.org/10.1364/ao.XX.XXXXXX}}
	
	\begin{abstract}
	In-line digital holography is a valuable tool for sizing, locating and tracking micro- or nano-objects in a volume. When a parametric imaging model is available, Inverse Problems approaches provide a straightforward estimate of the object parameters by fitting data with the model, thereby allowing accurate reconstruction. As recently proposed and demonstrated, combining pixel super-resolution techniques with Inverse Problems approaches improves the estimation of particle size and 3D-position. Here we demonstrate the accurate tracking of colloidal particles in Brownian motion. Particle size and 3D-position are jointly optimized from video holograms acquired with a digital holographic microscopy set up based on a ``low-end'' microscope objective ($\times 20$, $\rm NA\ 0.5$). Exploiting information redundancy makes it possible to characterize particles with a standard deviation of 15 nm in size and a theoretical resolution of 2 x 2 x 5 nm$^3$ for position under additive white Gaussian noise assumption.
	\end{abstract}
	
	\setboolean{displaycopyright}{true}
	
\begin{document}
	
	\maketitle
	\thispagestyle{fancy}
	\ifthenelse{\boolean{shortarticle}}{\abscontent}{}
	
	\section{Introduction}
	Accurate 3D tracking of micro- or nano-objects is instructive in a wide range of scientific domains (including biomedical imaging, fluid mechanics, microfluidics). Digital holographic microscopy (DHM) is well suited for these studies as it allows 3D quantitative imaging with temporal resolution. These properties have led to the development of video holographic microscopy, which has been shown to be a reliable and accurate tool for tracking colloidal particles~\cite{Lee2007}. Video holographic microscopy has also been shown to be useful for 3D tracking of micro-~\cite{Lee2007,Cheong2010,Dettmer2014} and nano- particles~\cite{Verpillat2011,Martinez-Marrades2014} or rods~\cite{Cheong2010rod} in Brownian motion, as well as for the characterization of particle interactions~\cite{Fung2011}. In these studies, tracking accuracy down to $3\times 3\times 10\ \rm nm^3$ for backpropagation reconstruction methods~\cite{Martinez-Marrades2014}, and $1\times 1\times 10\ \rm nm^3$ for Lorenz-Mie fitting~\cite{Lee2007} have been demonstrated. Deconvolution strategies can be used to track particles in three dimensions with similar axial and lateral resolutions, in that case leading to nanometer resolution in the three spatial dimensions~\cite{Dixon2011}. However, in these studies, only high-end microscope objectives (high magnification, high numerical aperture) or dark-field configuration have been considered, thus hindering the generalization of these results to more cost effective devices~\cite{Bishara2010,Mudanyali2013}.
	
	``Inverse Problems'' (IP) approaches have been shown to be more accurate than classical reconstruction methods~\cite{SoulezDenisFounier2007,SoulezDenisThiebaut2007,Bourquard2013}, and also to be optimal in certain experimental configurations~\cite{Mortensen2010,Fournier2010}. Instead of transforming acquired data with light backpropagation algorithms, these methods aim to find values of the imaging model parameters that maximize the likelihood with acquired data. The combination of IP approaches with ``pixel super-resolution (SR)'' which exploits the redundant information in sequences of video holograms has recently been shown to improve object parameter estimates, with a gain in variance at least proportional to the number of frames per reconstructed sequence~\cite{VerrierFournier2015}. This result thus paves the way for high-resolution 3D tracking of objects (e.g. colloidal particles), in a more affordable experimental set up.

	\begin{figure}[h]
	\centering
	\includegraphics[width = 7.5 cm]{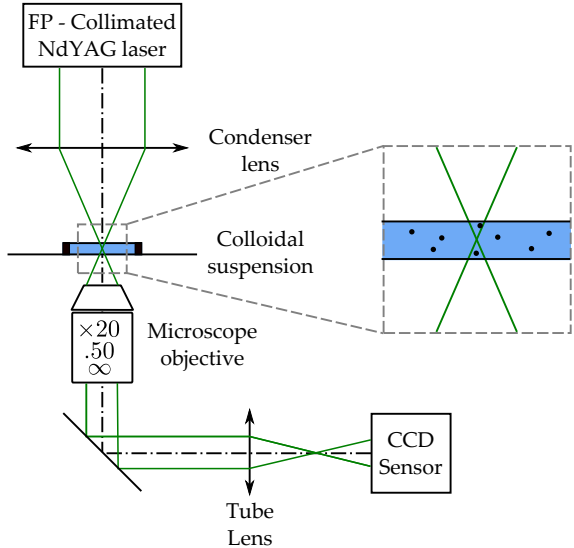}
	\caption{(Color online) Experimental set up for hologram acquisition.}\label{Fig:Setup}
	\end{figure}

	We take advantage of the recently proposed joint IP reconstruction approach to accurately track colloidal particles in Brownian motion. We first describe the reconstruction procedure used for accurate particle tracking from video hologram sequences is described. We then discuss the accuracy of the proposed procedure for the estimation of both 3D position and particle size, whose results are comparable to state of the art with a more affordable experimental configuration. Finally, obtained trajectory statistics are compared with Brownian motion theory statistics. 
	\begin{figure}[b]
	\centering
	\includegraphics[width = 8.2 cm]{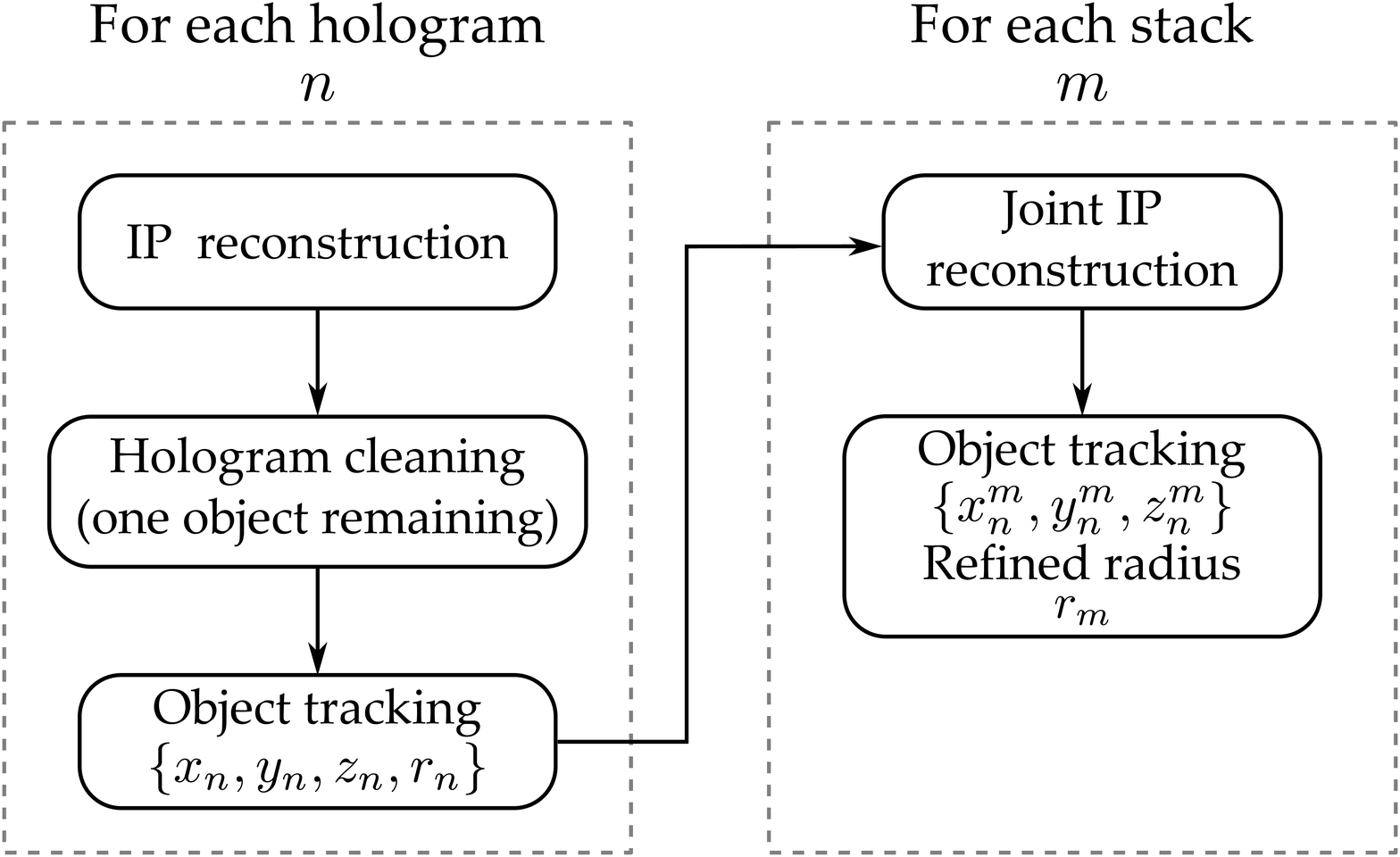}
	\caption{Synoptics of the joint IP reconstruction algorithm.}\label{Fig:Synoptics}
	\end{figure}

	\section{Data acquisition and processing}
	\subsection{Experimental set up}
	The experimental set up for hologram acquisition is depicted in Fig. \ref{Fig:Setup}. It consists of an inverted microscope (Olympus IX71 \textregistered) modified to operate in an in-line holographic configuration. Light emitted by a collimated fiber-pigtailed $\lambda=532\  \rm nm$ Nd-YAG laser (Coherent Verdi \textregistered~\ operating at 80 mW), is condensed on the sample, which consists of a colloidal suspension of $2r=1\ \mu\rm m$ in diameter polystyrene beads (standard deviation less than 0.1 $\mu$m according to the specifications) in permuted water (FLUKA 89904 \textregistered) maintained between the microscope slide and the cover slide with a 1 mm thick rubber spacer. Interference between the light scattered by polystyrene beads and the reference beam are collected using a $\times 20,\ \rm NA=0.5$ microscope objective (Olympus UPlanFLN \textregistered) corrected at infinity. Finally, the holograms are recorded on a 12 bit CCD sensor (Basler acA1600-20 um \textregistered, $1626\times1236$ pixels, with 4.4 $\mu\rm m$ pitch). Effective magnification of our experimental arrangement was estimated to be 21.12, leading to pixel pitches in the object half-space of $\delta_{\rm pix}^{\rm obj}=208.3\ \rm nm$, and an imaging field of $340\times 260\ \mu\rm m^2$. 
	\begin{figure}[t]
	\centering
	\includegraphics[width = 7.3 cm]{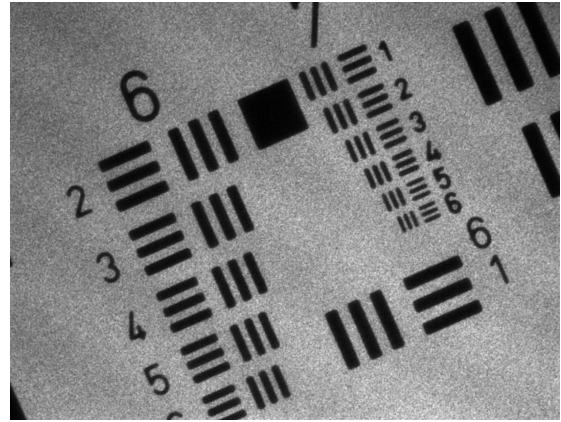}
	\caption{White light image of a USAF resolution target used to calibrate magnification.}\label{Fig:USAF}
	\end{figure}
	Estimation of the imaging field extension was performed by imaging an USAF resolution target under white light illumination (see Fig. \ref{Fig:USAF}). Images were recorded with a $\omega_{\rm CCD}/(2\pi)=20$ Hz framerate and $\tau_{\rm exp}=50 \times 10^{-6}\ \rm s$ exposure time.  For object tracking, a sequence of 1000 holograms consisting of a 50 s, 3D Brownian trajectory was recorded. Objects located between $z_{\rm min}=50\ \mu\rm m$ and $z_{\rm max}=70\ \mu\rm m$ were considered. This range of distances denotes the position of the analyzed objects compared to the image of the CCD sensor through the microscope objective. Within the specified distance range, the assumption $z\gg 4r^2/\lambda$ holds. Under this assumption, the diffracting particles can be considered as opaque, and the classical Huygens-Fresnel diffraction integral finds an analytic solution~\cite{Tyler1976}. The intensity $I_z\left(x,y\right)$ recorded for a spherical opaque object located at a distance $z$ from the imaging sensor is thus given by
	\begin{multline}\label{Eq:Thompson}
	I_z\left(x,y\right) \propto 1 - \frac{1}{\lambda z}\mathcal{F}_{\frac{x}{\lambda z},\frac{y}{\lambda z}}\left\{\vartheta\left(\xi,\eta\right)\right\}\sin\left[\frac{\pi}{\lambda z}\left(x^2+y^2\right)\right]\\
	+\left[\frac{1}{\lambda z}\mathcal{F}_{\frac{x}{\lambda z},\frac{y}{\lambda z}}\left\{\vartheta\left(\xi,\eta\right)\right\}\right]^2,
	\end{multline}
	where $\vartheta$ is the 2D aperture function of the particle under investigation, which is unity within the aperture and zero elsewhere. 
	Considering an opaque diffraction particle, the Fourier transform of $\vartheta$ is 
	\begin{equation}\label{Eq:Bessel}
	\mathcal{F}_{\frac{x}{\lambda z},\frac{y}{\lambda z}}\left\{\vartheta\left(\xi,\eta\right)\right\}=\pi r^2\left(\frac{\lambda z}{2\pi r \sqrt{x^2+y^2}}\right) J_1\left(\frac{2\pi r \sqrt{x^2+y^2}}{\lambda z}\right),
	\end{equation}
	with $J_1$ the Bessel function of the first kind.
	The main advantage of this image formation model is that it is analytic and depends on only the 3D positions and radius of the particle. We therefore use it in the remainder of the article.
	\begin{figure}[b]
	\centering
	\includegraphics[width = 8.4 cm]{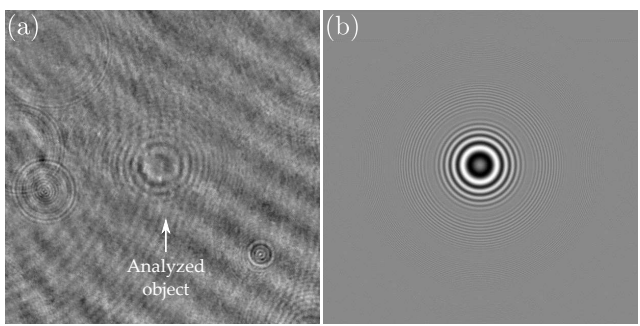}
	\caption{(a) Hologram of 1 $\mu$m diameter latex beads. The tracked particle is indicated by a white arrow. (b) Imaging model extracted from IP reconstruction (a) (See Media 1 and 2). Holograms were cropped to $512\times 512$ pixels around the tracked particle for the purpose of illustration.}\label{Fig:CleaningHologram}
	\end{figure}

	\subsection{Joint ``Inverse Problems'' reconstruction}
	Hologram sequences were processed using a recently proposed IP reconstruction scheme based on joint estimation of image model parameters that are experimentally invariant from one frame to the next~\cite{VerrierFournier2015}. In the present study, the algorithm was adapted to 3D tracking (see Fig. \ref{Fig:Synoptics}): joint estimation was performed on the radius of the object, which, in our set up is the only invariant parameter on the image sequence.
	
	The algorithm comprises two main steps.
	\begin{enumerate}
	\item First, each recorded hologram is processed using a classical IP reconstruction algorithm~\cite{SoulezDenisFounier2007,SoulezDenisThiebaut2007}. This algorithm identifies the parameters of the imaging model that best fit (in the least square sense) the acquired data. It consists of three steps:
	\begin{enumerate}
	\item A global, or coarse detection step, which find the best-matching model in the discretized parameter space $\left\{x,y,z,r\right\}$.  
	\item A local detection step, in which the coarsely estimated values are continuously optimized using a gradient decent algorithm.
	\item A cleaning step, in which the detected object is removed from the original data in order to increase the hologram signal to noise ratio (SNR).
	\end{enumerate}
	This procedure is iterated, for each hologram of the sequence, until no more particles are detected.
	
	As multiple objects are detected on each hologram in the sequence, all the detected particles except one are cleaned (i.e. based on an additive model assumption all diffraction patterns of detected objects are subtracted from the data). This step increases the SNR of the remaining object for accurate tracking and sizing. For the remaining object, we thus obtain the position and size parameters $\left\{x_n,y_n,z_n,r_n\right\}_{n=1..N_t}$ for the $N_t=1000$ recorded holograms. 
	This step is illustrated in Fig. \ref{Fig:CleaningHologram}, which shows holograms cropped to $512\times 512$ pixels around the tracked object. It should be noted that the holograms were cropped for the purpose of illustration, and that hologram processing was performed on the whole image size (i.e. $1626\times 1236$ pixels). The object, indicated by a white arrow (Fig. \ref{Fig:CleaningHologram}(a)) was tracked using an imaging model as depicted in Fig. \ref{Fig:CleaningHologram}(b) (see Media 1 and 2).
	
	\item In the second step a joint IP reconstruction was performed on $M=100$ image stacks, each composed of $N=10$ holograms, built from the $N_t=N\times M=1000$ recorded holograms. For each image stack $m$, the particle position parameters $\left\{x_n,y_n,z_n\right\}_{n=(m-1)N+1 ..mN}$, and the averaged radius over $N$ images $\langle r \rangle ^m_{indiv}=\frac{1}{N} \sum_{n=(m-1)N+1}^{mN} r_n$ were used as a starting point for the joint IP reconstruction. The particle radius $r^m_{joint}$ was estimated jointly, whereas $\left\{x_n^m,y_n^m,z_n^m\right\}_{n=1..N}$ were optimized individually.
	\end{enumerate}
	\begin{figure}[t]
	\centering
	\includegraphics[width = 8.4 cm]{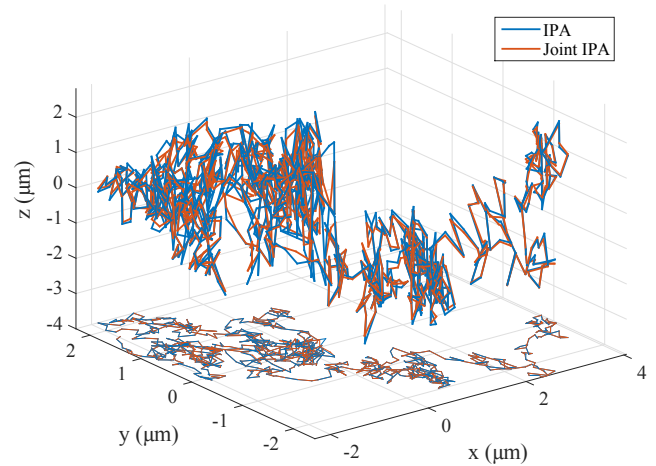}
	\caption{(Color online) Trajectory extracted from the IP reconstruction (dark blue) and the joint IP algorithm (dark red) of the hologram sequence in Fig. \ref{Fig:CleaningHologram}. For the sake of readability, only projection in the $xy$ plane is proposed .}\label{Fig:BrownianMotionIPA}
	\end{figure}
	The resulting trajectories are illustrated in Fig. \ref{Fig:BrownianMotionIPA} for the classical IP approach (corresponding to the first step in dark blue in Fig. \ref{Fig:Synoptics}) and the joint IP approach (corresponding to the complete algorithm in dark red in Fig. \ref{Fig:BrownianMotionIPA}). As can be figured out from the $xy$ plane projection, the 3D positions of the tracked object have been slightly modified thanks to the joint IP reconstruction. This point will be discussed in more detail in the following section.
	
	\section{Reconstruction accuracy}
	The ability of the IP reconstruction to track colloidal suspension in three dimensions has already been demonstrated in the literature. To prove the benefits of our processing scheme, the accuracy on the estimation of particle position parameters  $\left(x,y,z\right)$, and size $r$ is estimated in this section. It should be noted that, due to the small correlation between the imaging model parameters in our experiment, the improvement in accuracy provided by the joint IP reconstruction scheme is only significant for a parameter that remains constant along the hologram sequence (i.e. the particle radius)~\cite{Mortensen2010}. The estimation of the position parameters $\left(x,y,z\right)$ is consequently not changed. Since, $r$ is assumed to be constant along the hologram stack, its estimation accuracy is improved by our approach. 
	
	\subsection{Accuracy of the 3D localization: a Monte-Carlo approach}\label{Sec:MonteCarlo}
\begin{figure}[h]
	\centering
	\includegraphics[width = 8cm]{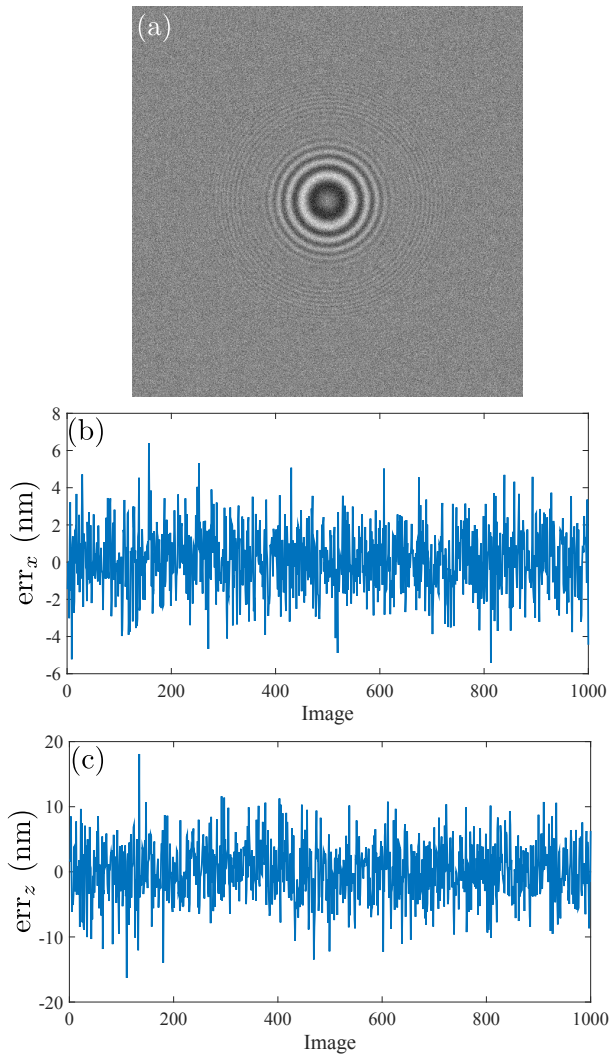}
	\caption{(Color online) Monte-Carlo estimation of the accuracy of the reconstruction. (a) Simulated hologram with a SNR similar to that of the experimental video hologram. The image has been cropped to 512 $\times$ 512 pixels for the purpose of illustration. Comparison between Monte-Carlo simulated and reconstructed values for (b) $x$, (c) $z$ coordinates.}\label{Fig:MonteCarlo}
	\end{figure}

	As the tracked colloidal particle is in Brownian motion, we cannot compare the estimated values of position parameters $\left\{x_n,y_n,z_n\right\}$ with ground truth values. To get round this problem, the accuracy of the estimation of the 3D positions of the tracked object was estimated with a Monte-Carlo simulation in which 1000 synthetic holograms of a spherical opaque object are generated. The 1000 $(x,y,z)$ positions are considered randomly around the mean experimental positions, while the object radius is assumed to be equal to its mean experimental value. White Gaussian noise is added to the simulation. The SNR (ratio of the magnitude of the signal to the standard deviation of the noise) is set at 5, which approximately corresponds to the SNR of the experimental holograms. An example of a Monte-Carlo simulated hologram, cropped to 512 $\times$ 512 pixels for the purpose of illustration, is shown in Fig. \ref{Fig:MonteCarlo}(a). 
	The simulated holograms were reconstructed using the joint IP approach, and error between Monte-Carlo simulated values $\left\{x_n,y_n,z_n\right\}^{\rm MC}$ and the estimated values $\left\{x_n,y_n,z_n\right\}^{\rm est}$ were computed. The results are shown in Fig. \ref{Fig:MonteCarlo}, in which errors between simulated values and estimated values, denoted ${\rm err_x}$, and ${\rm err_z}$, are plotted against the image index of the video hologram sequence. The three dimensional accuracy of our estimation is given by the standard deviation of the error distributions in Fig. \ref{Fig:MonteCarlo}. We obtained a localization accuracy of $\sigma_x\times\sigma_y\times\sigma_z=2\times 2\times 5\ \rm nm^3$. This accuracy is comparable with that of previous studies~\cite{Lee2007,Verpillat2011,Martinez-Marrades2014} performed with high-end microscope objectives, which result in a smaller field of view than our set up. In our case, fitting on the signal high spatial frequencies, allowed us to achieve a lateral accuracy of one hundredth of a pixel. The main purpose of the proposed Monte-Carlo simulations was to assess the accuracy of our 3D localization scheme. Note that these simulations also make it possible to evaluate the maximum achievable accuracy $\sigma_{r_{\rm MC}}$ of the particle size  estimation. Considering the above mentioned simulation parameters, the achievable accuracy is $\sigma_{r_{\rm MC}}=8.5\ \rm nm$.
	\begin{figure}[t]
	\centering
	\includegraphics[width = 8 cm]{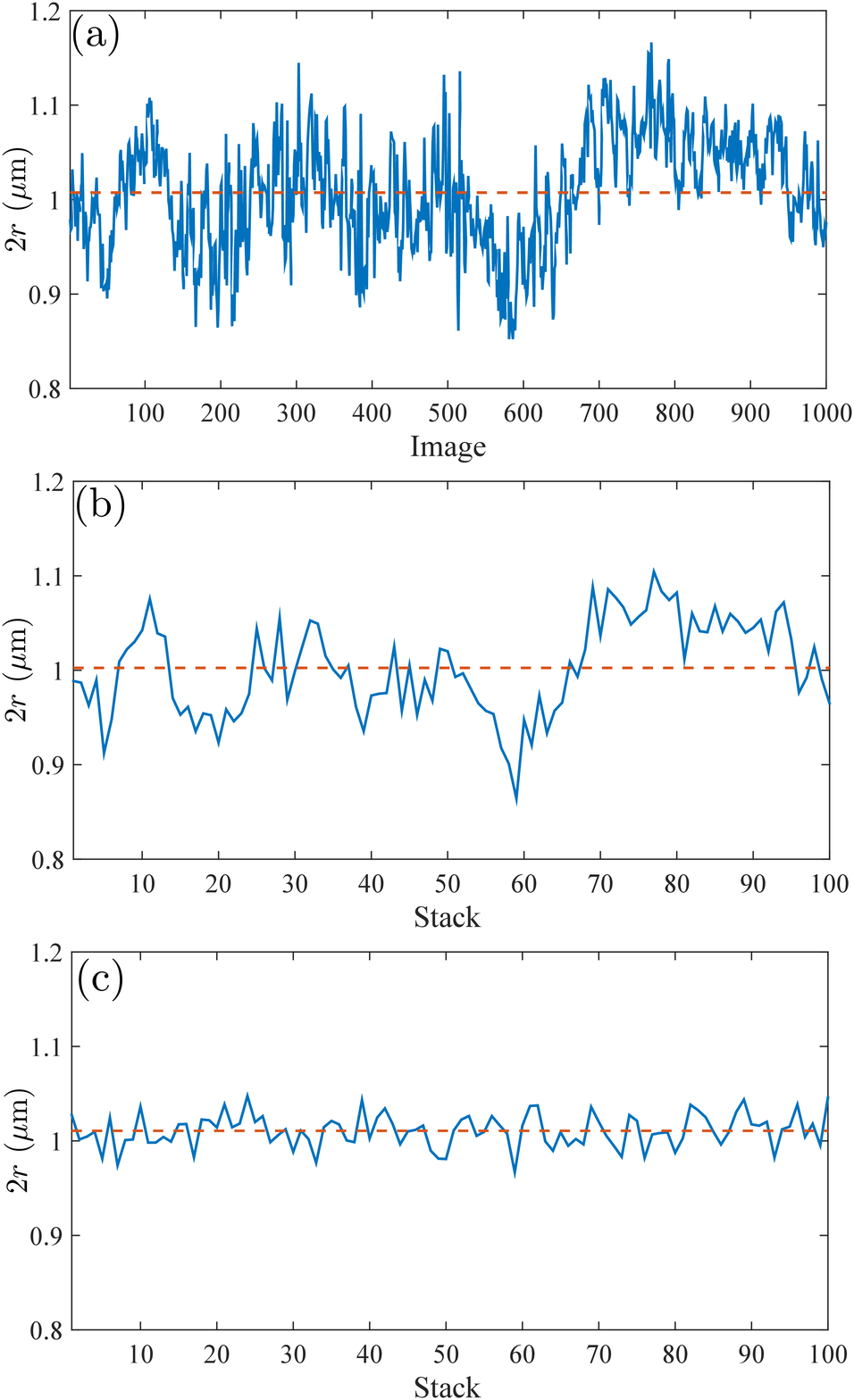}
	\caption{(Color online) Evolution of the estimated particle radius for (a) classical IP reconstruction $\left\{r_n\right\}_{n=1..1000}$; (b) joint IP reconstruction, with 10 consecutive holograms per image sequence $\left\{r^m\right\}_{m=1..100}$; (c) joint IP reconstruction, with 10 randomly chosen holograms per image sequence $\left\{r^{m}_{\rm rand}\right\}_{m=1..100}$. Dashed dark red line corresponds to the estimated average value of the particle radius.}\label{Fig:DispersionRayon}
	\end{figure}

	Note that these results are related to the accuracy of our IP reconstruction algorithm for the tracking of an individual object under the assumption of an additive white Gaussian detection noise. The above analysis therefore provides insights into the ultimately achievable accuracy. Nevertheless, as can be seen in Fig. \ref{Fig:CleaningHologram} and Media 1, several objects are in the field of view, leading to an inappropriate Gaussian white noise assumption. A more objective assessment of the tracking accuracy can be achieved by analyzing the acquired data statistics.
	
	\subsection{Accuracy of the 3D localization: analysis of experimental data statistics}
	Brownian particle positions cannot be predicted. Nevertheless, the fact that Brownian trajectories rely on known statistics can be advantageously used for the characterization of the tracking accuracy. This analysis relies on the results proposed by the authors of~\cite{Krishnatreya2014}. As predicted by Einstein~\cite{Einstein1956} and Smoluchowski~\cite{Smoluchowski1906}, the mean-square displacement of a particle diffusing in a liquid during the time interval $\Delta t$ is
	\begin{eqnarray}\label{Eq:MeanSqDisp}
	\Delta r^2\left(\Delta t\right) & = & \langle\left|\mathbf{r}\left(t+\Delta t\right)-\mathbf{r}\left(t\right)\right|^2\rangle\nonumber\\
	& = & 2dD\Delta t + \left(v\Delta t\right)^2,
	\end{eqnarray}
	where $d$ is the dimensionality of the concerned trajectory, $D$ is the diffusion coefficient of the particles in the surrounding liquid, $\mathbf{r}\left(t\right)=\left(x(t),y(t),z(t)\right)$ is the coordinate vector of the concerned trajectory, and $v$ is the diffusing particle velocity in the surrounding liquid.
	
	Using our $N$ experimental holograms, we can build a relation similar to that of Eq. (\ref{Eq:MeanSqDisp}) with different time intervals $\Delta t = s\times 2\pi/\omega_{\rm CCD}$~\cite{Krishnatreya2014}
	\begin{equation}\label{Eq:MeanSqDispExp}
	\Delta r_s^2=\left(\left\lfloor\frac{N}{s}\right\rfloor\right)^{-1}\sum_{k=1}^{\lfloor N/s \rfloor}\left|\mathbf{r}_{\left(n+1\right)s}-\mathbf{r}_{\left(n\right)s}\right|^2,
	\end{equation}
	where $s$ is the integer number of time steps considered for estimation of the mean-squared displacement, and $\lfloor \frac{N}{s} \rfloor$ is the entire part of the ratio $N/s$.
	\begin{figure}[t]
	\centering
	\includegraphics[width = 8.4 cm]{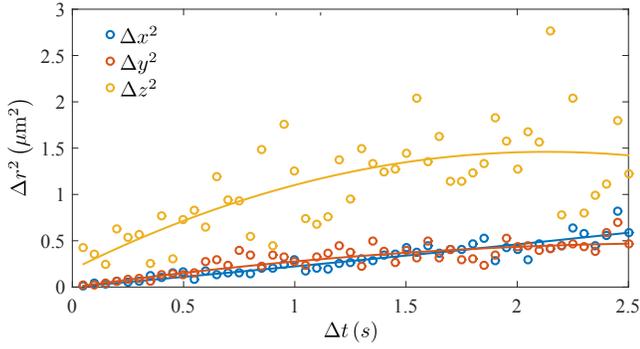}
	\caption{(Color online) Evolution of the mean-squared displacement as a function of the observation time $\Delta t$. Blue, red and yellow circles are respectively obtained by computing Eq. (\ref{Eq:MeanSqDispExp}) for $x$, $y$, and $z$ coordinates. Solid lines correspond to the fit of the experimental data according to Eq. (\ref{Eq:MeanSqDispOffset}).} \label{Fig:MSD}
	\end{figure}
	To account for measurement variance, the authors of~\cite{Krishnatreya2014} have proposed a modification of Eq. (\ref{Eq:MeanSqDisp}) 
	\begin{equation}\label{Eq:MeanSqDispOffset}
	\Delta r^2(\Delta t)=2d\varepsilon^2+2dD\Delta t+v^2\Delta t^2,
	\end{equation}
	where $2d\varepsilon^2$ is the measurement related offset.
	Therefore, fitting the mean-squared displacement obtained from the experimental displacement data (Eq. (\ref{Eq:MeanSqDispExp})) with Eq. (\ref{Eq:MeanSqDispOffset}) makes it possible to estimate the accuracy of Brownian tracking. From these fits (see Fig. \ref{Fig:MSD}), one can obtain the tracking errors for each coordinate of $\varepsilon_x = 17\pm40\ \rm nm$, $\varepsilon_y = 18\pm40\ \rm nm$, and $\varepsilon_z = 65\pm60\ \rm nm$. Values of the errors as well as the uncertainty were obtained considering a weighted non-linear regression scheme: the values of the errors are given by the value at $\Delta t= 0$ of the fits, and the uncertainties are determined through the fitting response with a confidence interval of $95\ \%$. Fig. \ref{Fig:MSD} shows that the accuracy of the axial displacement $\Delta z^2$ is poorer than that of lateral displacements $\Delta {x,y}^2$. This can be explained by the numerical aperture of the microscope objective used ($\rm NA=0.5$). Nevertheless, these values are in agreement with the values obtained in~\cite{Krishnatreya2014} considering the magnification and numerical aperture of the imaging objective used in our study ($\times 20\ \rm NA=0.5$ instead of $\times 100\ \rm NA=1.4$ in Ref. \cite{Krishnatreya2014}).
	
	Although they are statistically consistent with the theoretical errors estimated in Sec. \ref{Sec:MonteCarlo} (i.e. $\sigma_x\times\sigma_y\times\sigma_z=2\times 2\times 5\ \rm nm^3$), the experimental errors are noticeably higher. This can be partially explained by the fact that the additive white Gaussian detection noise hypothesis was not verified. As a matter of fact, several objects can be found in the field of view, leading to a correlated noise as well as to interaction effects that are not accounted for in our statistical analysis.
	
	\subsection{Accuracy on the colloidal suspension size}\label{Sec:AccuracyRadius}
	Experimental estimation of the colloidal particle size also benefits from the use of a joint IP reconstruction approach as illustrated in Fig. \ref{Fig:DispersionRayon}. Fig. \ref{Fig:DispersionRayon}(a) shows the evolution of the estimated radius obtained  by reconstructing the holograms using a classical IP approach to the whole image sequence of 1000 holograms. The radius estimates have a mean value equal to $\langle 2r\rangle_{\rm indiv}=1.007\ \mu\rm m$ which is compatible with the vendor's colloidal suspension specifications.
	The figure also shows that the estimate is noisy. The standard deviation is equal to $\sigma_{r_{\rm indiv}}=58\ \rm nm$. It should be noted that part of this dispersion is the result of the bias, which varies in the sequence.       
	Figure \ref{Fig:DispersionRayon}(b) shows the evolution of the estimated radius obtained with a joint estimation of the radius in the stacks. Its mean value is similar to the previous one i.e. $\langle 2r\rangle_{\rm joint}=1.002\ \mu\rm m$. The dispersion of the measurements is slightly lower, but a non constant bias remains leading to a standard deviation of $\sigma_{r_{\rm joint}}=49\ \rm nm$.
	As can be seen in Movie 1, particles that falls outside of our detection range (and are therefore not cleaned by our IP reconstruction procedure) are superimposed on the detected object pattern. This effect is likely to contribute to the bias in radius estimation. 
	
	To reduce this time correlated noise, the joint estimation can be performed on stacks of holograms randomly selected in the whole sequence. Results obtained using stack of images built with randomly selected holograms are plotted in Fig. \ref{Fig:DispersionRayon}(c). As expected, the bias in the radius estimation was successfully removed. This confirms the fact that bias in the estimated radius is closely linked to correlated noise originating from the presence of unwanted particles in the field of view. From these measurements, the colloidal suspension size was estimated to be $\langle 2r\rangle_{\rm rand}=1.005\ \mu\rm m$, with a standard deviation of $\sigma_{r_{\rm rand}} = 15\ \rm nm$. Nevertheless, it should be noted that the obtained accuracy of the radius estimation is 50 $\%$ higher than the one proposed in \cite{Lee2007}, which can be explained by the fact that, in our study, we used smaller numerical aperture objectives ($\rm NA=0.5$ than the $\rm NA=1.4$ in \cite{Lee2007}). 
	
	Standard deviations on the radius estimation for both classical ($\sigma_{r_{\rm indiv}}$) and randomized joint IP reconstruction ($\sigma_{r_{\rm rand}}$) provide useful information for our colloidal particle as the closeness of the measured radius to the average of the radii distribution is assessed. As we assume that the radius of the particles remains constant over time, it is also relevant to estimate the standard error of the mean (i.e. the standard deviation of the sampling distribution of the sample mean) for both methods. Considering a distribution $x$, the standard error of the mean $\sigma_{\bar{x}}$ is
	\begin{equation}\label{Eq:SEmean}
	\sigma_{\bar{x}}=\frac{\sigma_x}{\sqrt{N_{\rm s}}},
	\end{equation}
	where $\sigma_x$ is the standard deviation of the distribution, and $N_{\rm s}$ is the number of independent observations. Therefore, using Eq. (\ref{Eq:SEmean}), the standard error of the mean radius for classical IP reconstruction is $\sigma_{\bar{r}_{\rm indiv}}=1.8\ \rm nm$, while it drops to $\sigma_{\bar{r}_{\rm rand}}=1.5\ \rm nm$ for the randomized joint IP reconstruction. These errors were estimated for $N_{\rm s}=N_t$ (classical IP) and $N_{\rm s}=M$ (randomized joint IP) respectively.
	\begin{table}[htbp]
	\centering
	\caption{\bf Standard deviation and standard error of the mean of the radius estimation for both classical and randomized joint IP reconstruction}
	\begin{tabular}{ccc}
	\hline
	 & $\sigma_r\ \left(\rm nm\right)$ & $\sigma_{\bar{r}}\ \left(\rm nm\right)$ \\
	\hline
	Classical IP & $58$ & $1.8$ \\
	Randomized joint IP & $15$ & $1.5$ \\
	\hline
	Improvement & $280\ \%$ & $22\ \%$\\
	\hline
	\end{tabular}
	  \label{Tab:Results}
	\end{table}
	Results are summarized in Table \ref{Tab:Results}. It can be seen that our randomized joint IP greatly improves the estimated radius dispersion. There is less improvement in the standard error of the mean radius, but it nevertheless demonstrates the benefits of using a joint estimation of redundant parameters in sequences of holograms.
	
	\subsection{Comparison with Brownian motion theory}
	To prove the advantages of our approach in Brownian motion tracking, we calculated the theoretical average of the 3D displacement of a $2r=1\ \mu\rm m$ polystyrene bead in water. According to Stokes-Einstein theory~\cite{Einstein1956} the average squared-displacement is
	\begin{equation}\label{Eq:DisplacementTh}
	\langle \Delta r^2\rangle=6D\tau_{\rm obs},
	\end{equation}
	where $\tau_{\rm obs}$ is the ``length'' of observation (here $\tau_{\rm obs}=(2\pi)/\omega_{\rm CCD}=50 \times 10^{-3}\ \rm s$), and $D$ the diffusivity of the particle
	\begin{equation}\label{Eq:Diffusivity}
	D = k_B T\gamma^{-1},
	\end{equation}
	where $\gamma = 6\pi\mu r_p$, is the drag coefficient of the particle. The trajectory shown in Fig. \ref{Fig:BrownianMotionIPA} was obtained at room temperature $T=300\ \rm K$, for particles in water of dynamic viscosity $\mu=0.89\times10^{-3}\ \rm kg.m^{-1}.s^{-1}$. The radius of the particle $r_p$ corresponds to the colloidal suspension radius given by the manufacturer's specifications. From Eq. (\ref{Eq:DisplacementTh}), one can assess the ``so-called'' averaged displacement $\Delta'_{\rm th}$ defined as
	\begin{equation}
	\Delta r'_{\rm th}=\sqrt{\langle \Delta r^2\rangle}=\sqrt{6D\tau_{\rm exp}}
	\end{equation}
	that we used to demonstrate the advantages of our approach. According to the experimental parameter, the theoretical displacement of the particle from one frame to the next is $\Delta'_{\rm th}\approx 0.271\ \mu\rm m$. The experimental averaged displacement $d'_{\rm exp}$ can be estimated from data considering
	\begin{equation}\label{Eq:DisplacementExp}
	\Delta r'_{\rm exp}=\sqrt{\langle \Delta x^2\rangle+\langle \Delta y^2\rangle+\langle \Delta z^2\rangle},
	\end{equation}
	where $\Delta i_{i=x,y,z}$ is the 3D displacement of the object between two frames. Using Eq. (\ref{Eq:DisplacementExp}), the experimental averaged displacement is $\Delta r'_{\rm exp}\approx 0.251\ \mu\rm m$, which is in good agreement with the Brownian motion theory.
	\begin{figure}[t]
	\centering
	\includegraphics[width = 8.4 cm]{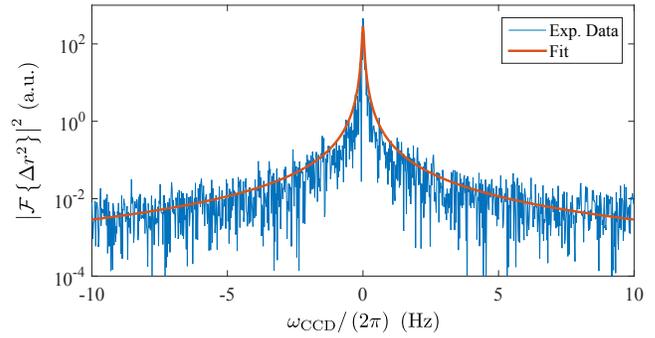}
	\caption{(Color online) Power spectral density of the experimental trajectory (blue) fitted with the theoretical Brownian motion power spectral density (orange). Results are plotted in a $y$-semilog scale.} \label{Fig:PSDBrown}
	\end{figure}

	This agreement was confirmed by estimating the power spectral density (PSD) of the reconstructed trajectory. Considering a mass-less particle, and according to the fluctuation-dissipation theorem, the PSD is
	\begin{equation}\label{Eq:PSDTh}
	\left|\mathcal{F}\left\{\Delta r^2\right\}\right|^2 = k_B T\gamma^{-1}\omega^{-2}.
	\end{equation}
	Therefore, fitting the PSD extracted from the experimental trajectory with Eq. (\ref{Eq:PSDTh}) made it possible to measure particle diffusivity $D$. Fig. \ref{Fig:PSDBrown} shows the results and the good agreement between the acquired data and the Brownian motion theory. For our trajectory, we obtained $D=0.2893\pm 0.0054\ \mu\rm m^2.s^{-1}$, which slightly overestimates the value calculated according to our experimental conditions ($D=0.2469\ \mu\rm m^2.s^{-1}$).
	
	\section{Conclusion}
	Colloidal particles in Brownian motion were tracked using a jointly optimized IP reconstruction algorithm. Using a cost effective experimental set up, and under assumption of ideal white Gaussian detection noise, the proposed approach makes it possible, to track in 3D $2r=1\ \mu\rm m$ polystyrene particles with a $2\times 2 \times 5\ \rm nm^3$ resolution and a 15 $\rm nm$ standard deviation on the particle radius estimation. Analysis of the experimental data made it possible to assess the tracking errors, that are noticeably higher than, -and statistically consistent with- the ideal theoretical values. The suitability of the approach was demonstrated by comparing the theoretical average Brownian motion displacement with experimentally observed displacement.  Bias in the experimental measurements was due to the presence of unwanted particles in the field of view, resulting in time-correlated background noise. To confirm this last point, the joint IP reconstruction algorithm was modified by randomly building image stacks. This resulted in debiasing of the colloidal suspension size estimation. It should be noted that the size of image stack used for joint estimation has an influence on the accuracy of the estimation. 
	
	It should also be noted that a simplified imaging model relying on four parameters (3D position and suspension size) was sufficient for this particular tracking study. Nevertheless, a more complete imaging model, such as Lorenz-Mie theory, can be considered for more severe experimental conditions. In that case, joint estimation of additional parameters such as the colloidal particle refractive index would be possible, keeping in mind that the estimation of all the parameter remaining constant during the image acquisition will benefit from our joint IP reconstruction scheme.
	
	The authors are grateful for the financial support provided by Lyon University through the Programs \emph{"Investissements d'Avenir"} (ANR-11-IDEX-0007) and LABEX PRIMES (ANR-11-LABX-0063), and also wishes to acknowledge R. Volk and L. Denis for fruitful discussions.


\begin{thebibliography}{}
\newcommand{\enquote}[1]{``#1''}

\end{thebibliography}


\begin{thebibliography}{10}
	
	\bibitem{Lee2007}
	S.~H. Lee, Y.~Roichman, G.~R. Yi, S.~H. Kim, S.~M. Yang, A.~van Blaaderen,
	  P.~van Oostrum, and D.~G. Grier, ``Characterizing and tracking single
	  colloidal particles with video holographic microscopy,'' Opt. Express
	  \textbf{15}, 18275--18282 (2007).
	
	\bibitem{Cheong2010}
	F.~C. Cheong, B.~J. Krishnatreya, and D.~G. Grier, ``Strategies for
	  three-dimensional particle tracking with holographic video microscopy,'' Opt.
	  Express \textbf{18}, 13563--13573 (2010).
	
	\bibitem{Dettmer2014}
	S.~L. Dettmer, U.~F. Keyser, and S.~Pagliara, ``Local characterization
	  of hindered brownian motion by using digital video microscopy and 3d particle
	  tracking,'' Rev. Sci. Instrum. \textbf{85}, 023708 (2014).
	
	\bibitem{Verpillat2011}
	F.~Verpillat, F.~Joud, P.~Desbiolles, and M.~Gross, ``Dark-field digital
	  holographic microscopy for 3d-tracking of gold nanoparticles,'' Opt. Express
	  \textbf{19}, 26044--26055 (2011).
	
	\bibitem{Martinez-Marrades2014}
	A.~Martinez-Marrades, J.-F. Rupprecht, M.~Gross, and G.~Tessier,
	  ``Stochastic 3D optical mapping by holographic localization of
	  brownian scatterers,'' Opt. Express \textbf{22}, 29191--29203 (2014).
	
	\bibitem{Cheong2010rod}
	F.~C. Cheong and D.~G. Grier, ``Rotational and translational diffusion
	  of copper oxide nanorods measured with holographic video microscopy,'' Opt.
	  Express \textbf{18}, 6555--6562 (2010).
	
	\bibitem{Fung2011}
	J.~Fung, K.~E. Martin, R.~W. Perry, D.~M. Kaz, R.~McGorty, and V.~N. Manoharan,
	  ``Measuring translational, rotational, and vibrational dynamics in
	  colloids with digital holographic microscopy,'' Opt. Express \textbf{19},
	  8051--8065 (2011).
	
	\bibitem{Dixon2011}
	L.~Dixon, F.~C. Cheong, and D.~G. Grier, ``Holographic deconvolution
	  microscopy for high-resolution particle tracking,'' Opt. Express \textbf{19},
	  16410--16417 (2011).
	
	\bibitem{Bishara2010}
	W.~Bishara, T.-W. Su, A.~F. Coskun, and A.~Ozcan, ``Lensfree on-chip
	  microscopy over a wide field-of-view using pixel super-resolution,'' Opt.
	  Express \textbf{18}, 11181--11191 (2010).
	
	\bibitem{Mudanyali2013}
	O.~Mudanyali, E.~McLeod, W.~Luo, A.~Greenbaum, A.~F. Coskun, Y.~Hennequin,
	  C.~P. Allier, and A.~Ozcan, ``Wide-field optical detection of
	  nanoparticles using on-chip microscopy and self-assembled nanolenses,'' Nat.
	  photonics \textbf{7}, 247--254 (2013).
	
	\bibitem{SoulezDenisFounier2007}
	F.~Soulez, L.~Denis, C.~Fournier, E. Thi\'{e}baut, and C.~Goepfert,
	  ``Inverse-problem approach for particle digital holography: accurate
	  location based on local optimization,'' J. Opt. Soc. Am. A \textbf{24},
	  1164--1171 (2007).
	
	\bibitem{SoulezDenisThiebaut2007}
	F.~Soulez, L.~Denis, E. Thi\'{e}baut, C.~Fournier, and C.~Goepfert,
	  ``Inverse problem approach in particle digital holography:
	  out-of-field particle detection made possible,'' J. Opt. Soc. Am. A
	  \textbf{24}, 3708--3716 (2007).
	  
	\bibitem{Bourquard2013}
	A. Bourquard, N. Pavillon, E. Bostan, C. Depeursinge, and M. Unser,
	  ``A practical inverse-problem approach to
	digital holographic reconstruction,'' Opt. Express
	  \textbf{21}, 3417--3433 (2013).  
	
	\bibitem{Mortensen2010}
	K.~I. Mortensen, L.~S. Churchman, J.~A. Spudich, and H.~Flyvbjerg,
	  ``Optimized localization analysis for single-molecule tracking and
	  super-resolution microscopy,'' Nat. Methods \textbf{7}, 377--381 (2010).
	
	\bibitem{Fournier2010}
	C.~Fournier, L.~Denis, and T.~Fournel, ``On the single point resolution
	  of on-axis digital holography,'' J. Opt. Soc. Am. A \textbf{27}, 1856--1862
	  (2010).
	
	\bibitem{VerrierFournier2015}
	N.~Verrier and C.~Fournier, ``Digital holography super-resolution for
	  accurate three-dimensional reconstruction of particle holograms,'' Opt. Lett.
	  \textbf{40}, 217--220 (2015).
	
	\bibitem{Tyler1976}
	G.~Tyler and B.~Thompson, ``Fraunhofer holography applied to particle
	  size analysis a reassessment,'' J. Mod. Opt. \textbf{23}, 685--700 (1976).
	
	\bibitem{Krishnatreya2014}
	B.~J.~Krishnatreya, A.~Colen-Landy, P.~Hasebe, B.~A.~Bell, J.~R.~Jones, A.~Sunda-Meya, and D.~G.~Grier, ``Measuring Boltzmann's constant through holographic video microscopy of a single colloidal particle,'' Am. J. Phys. \textbf{82}, 23--31 (2014).
	
	\bibitem{Einstein1956}
	A.~Einstein, \emph{Investigations on the Theory of the Brownian Movement}
	  (Courier Corporation, 1956).
	
	\bibitem{Smoluchowski1906}
	M.~von Smoluchowski, ``Zur kinetischen Theorie der Brownschen Molekularbewegung und der Suspensionen,''
	  Ann. der Phys. \textbf{326}, 756-780.
	
	\end{thebibliography}
	\end{document}